\documentclass[12pt,preprint]{aastex}         
\def\be{\begin{equation}}
\def\ee{\end{equation}}

\def\bb{\mbox{\boldmath $\beta $}}
\def\oo{\mbox{\boldmath $\Omega $}}

\def\lsim{\lower 2pt \hbox{$\, \buildrel {\scriptstyle <}\over
         {\scriptstyle \sim}\,$}}
\newcommand\gsim{\buildrel > \over \sim}
\begin{document}
\newcommand{\figureout}[2]{ \figcaption[#1]{#2} }       

\title{PAIR-STARVED PULSAR MAGNETOSPHERES}

\author{Alex G. Muslimov\altaffilmark{1,2} \& 
Alice K. Harding\altaffilmark{2}}   

\altaffiltext{1}{Universities Space Research Association, Columbia, MD 21044}

\altaffiltext{2}{Astrophysics Science Division,      
NASA/Goddard Space Flight Center, Greenbelt, MD 20771}
 

\begin{abstract}
We propose a simple analytic model for the innermost (within the light cylinder of canonical radius, $\sim c/\Omega $) structure of open-magnetic-field lines of a rotating neutron star (NS) with relativistic outflow of charged particles (electrons/positrons) and arbitrary angle between the NS spin and magnetic axes. We present the self-consistent solution of Maxwell's equations for the magnetic field and electric current in the pair-starved regime where the density of electron-positron plasma generated above the pulsar polar cap is not sufficient to completely screen the accelerating electric field and thus establish the  ${\bf E} \cdot {\bf B} = 0$  condition above the pair-formation front up to the very high altitudes within the light cylinder. The proposed model may provide a theoretical framework for developing the refined model of the global pair-starved pulsar magnetosphere.        
\end{abstract} 


\pagebreak
  
\section{INTRODUCTION}

It is believed that rotation-powered pulsars possess a magnetosphere in which charged particles are accelerated to relativistic Lorentz factors and generate a broad spectrum of pulsed emission (from radio, to IR, optical, X-ray and $\gamma $-ray).  The detailed physics of the global magnetosphere depends on various factors such as the efficiency of primary-particle (electron) ejection and acceleration near the neutron star (NS) surface, conditions for the occurrence and intensity of electron-positron pair creation, the spatial distribution of electromagnetic fields and currents, and particle energy loss mechanisms.  Although the vacuum magnetosphere model of Deutsch (1955) is, at present, the only available non-axisymmetric closed analytic model for the electromagnetic field of a rotating magnetic dipole, it is not an appropriate physical model of an active pulsar magnetosphere filled with charges and currents.  Historically, the development of more realistic models began with the idealized equation governing the structure of an axisymmetric pulsar magnetosphere (see e.g. Mestel 1973; Scharlemann \& Wagoner 1973; Michel 1973; Okamoto 1974; Mestel et al. 1979 and references therein). This equation, sometimes referred to as the pulsar equation, is the astrophysical counterpart of the force-free Grad-Shafranov equation (see the original publications by Grad [1967] \& Shafranov [1966]).  Further development was undertaken by Mestel (1999), Goodwin et al. (2004), Beskin et al. (1983), and Beskin, Kuznetsova \& Rafikov (1998), and significant progress has been achieved in the numerical solution (see Contopoulos et al. 1999; Mestel 1999; Spitkovsky 2006; Timokhin 2006, 2007) of the pulsar equation in the ideal-MHD (${\bf E} \cdot {\bf B} = 0$) and force-free (neglect of particle inertia and pressure) approximation (see e.g. Arons 2004 for a brief theoretical overview).  Although the force-free magnetosphere is probably a closer approximation to a real pulsar than the vacuum solution, it is still not a truly self-consistent model since the production of pair plasma requires particle acceleration and thus a break down of force-free conditions in some regions of the magnetosphere.  Furthermore, it will apply to the younger pulsars that can readily supply the requisite charge through copious electron-positron pair cascades.

A different approach to the theory of pulsar magnetospheres has been the  
simulations begun by Krause-Polstorff \& Michel (1985, see also Smith et al. 2001, Petri et al. 2002), who studied how an axisymmetric rotating NS surrounded by vacuum fills with charge.  They found that the evolution of a rotating, conducting sphere results in separated domes of charge over the magnetic poles and a torii of opposite charge in the equator, with vacuum between.
Such a configuration is devoid of currents and charge outflow and is thus a dead pulsar.  However, more
recent 3D plasma simulations (Spitkovsky \& Arons 2002; Biltzinger \& Thielheim 2004; and Spitkovsky 2004, 2006, 2008) found that the charge domes are subject to diocotron instability, allowing charge from disrupted domes to fill the vacuum regions, opening up the possibility for a charge-filled magnetosphere. 

Studies of polar cap acceleration and pair cascades (Harding \& Muslimov 2001, 2002, Harding, Muslimov \& Zhang 2002) have found that only the youngest 
third of the known pulsar population (those that can produce pairs through curvature radiation) is capable of producing
enough charges in pair cascades to completely screen the parallel electric field.  The bulk of pulsars
(those that produce pairs only through inverse-Compton radiation) cannot supply enough charge to
screen the $E_{\parallel}$ and will be ``pair-starved".  The magnetospheres of pair-starved pulsars
will thus lie in a very different regime from that of the force-free magnetospheres.
In our previous study (see Muslimov \& Harding 2005; MH05) we discussed the analytic solution pertaining to the pair-starved regime (Muslimov \& Harding 2004; MH04) of particle outflow in the innermost magnetosphere of a NS with arbitrary pulsar obliquity, and derived the explicit expressions for the magnetic field and corresponding electric currents. We assumed that the current of primary relativistic electrons was the only source of perturbation for the originally pure dipole magnetic configuration.  Our model implied therefore the occurrence of an accelerating electric field, ${\bf E}_{\parallel } \neq 0$, on open field lines. On the contrary, for the force-free and MHD models the electric current on open field lines is assumed to be very close to the Goldreich-Julian (GJ) current to ensure that ${\bf E}\cdot {\bf B} = 0$.  In this paper we propose a slightly different approach that lies between the MH05 and the force-free/MHD approximations described above, since it implies a self-consistent electric current for open field lines that is greater than the primary current but still smaller than the local GJ current, so that there is non-vanishing accelerating electric field along the open field lines. In this study we explore a fully analytic steady-state solution illustrating that the initially dipolar magnetic field undergoes growing topological change at increasingly higher altitudes by the self-consistent electric current that generates the 3D-monopole-type magnetic configuration (the corresponding magnetic field strength, $B$, scales as $1/r^2$). In our solution both the electric current and magnetic field are determined self-consistently, and it is assumed that near the polar cap surface the current matches the self-limited current calculated for the magnetic field determined by a pure dipole with small corrections from the 3D monopole.  

The paper is organized as follows. In \S ~2 we present a set of basic equations for the electromagnetic field that will be employed in our study. In \S ~3 we present our analytic solution for the magnetic field and currents. In \S ~4 we discuss the results of our study and summarize our main conclusions. 

\section{Basic Equations}

The very general equations describing the electromagnetic field of a rotating NS in the Lab (inertial) frame are the first couple,
\be
{\bf \nabla} \cdot {\bf B} = 0,
\label{divB}
\ee
\be
{\bf \nabla} \times {\bf E} = - {1\over c} {{\partial {\bf B}}\over {\partial t}},
\label{curlE}
\ee
and the second couple of Maxwell's equations,
\be
{\bf \nabla } \cdot {\bf E} = 4 \pi \rho ,
\label{divE}
\ee
\be
{\bf \nabla}\times {\bf B} = {1\over c} {{\partial {\bf E}}\over {\partial t}} + 
{{4 \pi}\over c} {\bf j},
\label{curlB}
\ee
where $\rho $ and $\bf j$ are the electric charge and current densities, respectively; and all physical quantities, such as $\bf B$, $\bf E$, $\bf j$, and $\rho $ are defined in the Lab frame.

We will be searching for the steady-state solution to equations (\ref{divB})-(\ref{curlB}) in the domain of open-magnetic-field lines. In the steady state that is assumed to exist, the time derivatives in equations (\ref{curlE}) and (\ref{curlB}) are determined by the proper set of special-relativistic transformations of the electromagnetic quantities in the case of rotational motion of open-field lines relative to the Lab frame. Moreover, it is important that in our model, even within the light cylinder, the open-magnetic-field lines may rotate slightly differentially  without being wound up (see \S 3 below).  To get the system of steady-state Maxwell's equations, we can use the following special-relativistic transformations (see MH05) of partial time derivatives between the Lab frame (subscript ``Lab") and the frame of reference corotating with the open-field lines (subscript ``corot"):
\be
\left\{ {{\partial {\bf B}}\over {\partial t}}  \right\} _{\rm corot} = \left\{ {{\partial {\bf B}}\over {\partial t}}  \right\} _{\rm Lab} - {\bf \nabla } \times ({\bf u}_{\rm rot} \times {\bf B}),
\label{dB/dt}
\ee
\be
\left\{ {{\partial {\bf E}}\over {\partial t}}  \right\} _{\rm corot} = \left\{ {{\partial {\bf E}}\over {\partial t}}  \right\} _{\rm Lab} - {\bf \nabla } \times ({\bf u}_{\rm rot} \times {\bf E}) + {\bf u}_{\rm rot}~{\bf \nabla \cdot E},
\label{dE/dt}
\ee
\be
\left\{ {{\partial \rho }\over {\partial t}}  \right\} _{\rm corot} = \left\{ {{\partial \rho }\over {\partial t}}  \right\} _{\rm Lab} + {\bf u}_{\rm rot} \cdot {\bf \nabla} \rho,
\label{drho/dt}
\ee
where ${\bf u}_{\rm rot}$ (= $\oo \times {\bf r}$, and $\Omega $ is the angular velocity that can, generally, be differential) is the linear rotational  velocity of the open-field lines.

We assume that in a steady state, the time derivatives in the LHS of equations (\ref{dB/dt})-(\ref{drho/dt}) vanish (i.e. the electromagnetic fields ``seen" by the observer corotating with the magnetic flux tube are stationary), so that Maxwell equations (\ref{curlE}) and (\ref{curlB}) can be rewritten in the following form
\be
\nabla \times {\bf E} = - \nabla \times ( {\bb }_{\rm rot} \times {\bf B}),
\label{curlE-2}
\ee
\be
\nabla \times {\bf B} = \nabla \times ( {\bb }_{\rm rot} \times {\bf E}) - 
{\bb}_{\rm rot} \nabla \cdot {\bf E} + {{4\pi}\over c}{\bf j}. 
\label{curlB-2}
\ee
where ${\bb}_{\rm rot} = {\bf u}_{\rm rot}/c$.

\noindent
Also, the charge continuity equation,
\be 
{{\partial \rho }\over {\partial t}} + \nabla \cdot {\bf j} = 0,
\label{drho/dt-2}
\ee
with the help of relationship (\ref{drho/dt}) takes the form
\be
{\bf u}_{\rm rot }\cdot \nabla \rho - \nabla \cdot {\bf j} = 0.
\label{drho/dt-3}
\ee
By combining equations (\ref{divE}) and (\ref{curlB-2}), we get
\be
\nabla \times \left( {\bf B} - {{{\bf u}_{\rm rot}}\over c} \times {\bf E} \right) = 
{{4\pi}\over c} ( {\bf j} - \rho {\bf u}_{\rm rot}). 
\label{curlB-3}
\ee
Thus, the steady-state solution for the domain of magnetosphere with open-field lines is determined by equations (\ref{divB}), (\ref{curlE-2}), (\ref{drho/dt-3}) and (\ref{curlB-3}). Note that, since $\nabla \cdot {\bf u}_{\rm rot} = 0$ and $\nabla \cdot (\rho {\bf u}_{\rm rot}) = {\bf u}_{\rm rot} \cdot \nabla \rho $, 
equation (\ref{drho/dt-3}) translates into
\be
\nabla \cdot ( {\bf j} - \rho {\bf u}_{\rm rot}) = 0.
\label{divj}
\ee   
We can ignore the second term in the LHS of equation (\ref{curlB-3}) that is of order of $(u_{\rm rot}/c)^2 \ll 1$. To complete the formulation we need to specify the current density $\bf j$. We assume that 
\be
{\bf j} = {\bf j}^{\rm (1)} + {\bf j}_{\rm rot} + {\bf j}_{_{{\bf E}\times {\bf B}}},
\label{jtot}
\ee
where ${\bf j}^{\rm (1)}$ is the density of the self-consistent electric current supporting the magnetic field ${\bf B}^{(1)}$ (see formula [\ref{B1-1}] below) , ${\bf j}_{\rm rot}= \rho {\bf u}_{\rm rot}$ is the density of the electric current generated by the bulk rotational motion of charges, and ${\bf j}_{_{{\bf E}\times {\bf B}}}$ is the density of the ${\bf E}\times {\bf B}$-drift current. In our previous paper (see MH05) we discussed the situation where the current ${\bf j}^{\rm (1)}$ was solely determined by the current of primary electrons in the dipole magnetic field. In this paper the current density ${\bf j}^{\rm (1)}$ matches the current density of primary electrons in the dipole field only at very small altitudes and provides significant distortion of initially dipole field at higher altitudes within the LC. Hence, in the present study, we may justifiably ignore much smaller contributions to the current $\bf j$ produced by the second-order correction to the primary electron current and by the ${\bf E}\times {\bf B}$ current, respectively, (see formulae [60]-[62] and [76]-[78] of MH05) in the RHS of equation (\ref{jtot}). Because of the linearity of equation (\ref{curlB-3}) the corrections to ${\bf B}^{\rm d}$ calculated in MH05 can be added to the final formulae for $\bf B$ (see equations [\ref{Br}] - [\ref{Bphi}] below).

As in our previous study (see MH05), we shall use the magnetic spherical polar coordinates ($x=r/R_{\rm lc}$, $\theta $, $\phi $; where $R_{\rm lc} = c/\Omega $) with the polar axis along the magnetic moment of a NS.  The magnetic coordinates are appropriate for modeling the magnetic field structure in the vicinity of the NS (at $x \ll 1$) with any obliquity. However,  the magnetic coordinates are inconvenient in describing the effects of rotation on the global magnetosphere, simply because the rotation can break the symmetry with respect to the magnetic axis at $x \gsim 1$ and for arbitrary obliquity.  Intuitively, one may expect that, at higher altitudes, the magnetic coordinates get transformed into the spherical coordinates with the $z-$ axis along the NS rotation axis that determines the global symmetry.  Indeed, we can observe that (see e.g. formulae [73]-[75] in MH05), in magnetic coordinates, the effects of rotation (besides making the corresponding terms proportional to $\Omega $ and having certain radial dependence) enter formulae for the magnetic field structure in an elegant way, through a single function, $cosine$ of the angle between the NS rotation axis and radius-vector of a given point (in terms of magnetic polar coordinates),
\be
s = \cos \chi \cos \theta + \sin \chi \sin \theta \cos \phi ,
\label{s}
\ee
and its derivatives over $\theta $ and $\phi $. Here, according to our notation,  $\chi $ is the NS obliquity angle. Note that $s = 1$ determines the symmetry axis in magnetic coordinates which is the NS rotation axis.  In addition, the function $s$ is a ``toroidal potential" for the rotational linear velocity, ${\bb}_{\rm rot} = {\bf u}_{\rm rot}/c = (\Omega r^2 /c) (\nabla s \times {\bf n})$, where ${\bf n}$ is the unit vector in radial direction.  We therefore choose to express our solution for ${\bf B}^{(1)}$, which is in magnetic spherical polar coordinates, in terms of $s$.  
 
\noindent 
In our calculation the following couple of useful formulae will be employed,
\be
\left({{\partial s }\over {\partial \theta }}\right) ^2 + {1\over {\sin ^2 \theta }}\left( {{\partial s}\over {\partial \phi }}\right) ^2 = 1-s^2,
\label{sder1}
\ee
and
\be
{1\over {\sin \theta }} {{\partial }\over {\partial \theta }}\left(  \sin \theta {{\partial s}\over {\partial \theta }} \right)  + 
{1\over {\sin ^2 \theta }} {{\partial ^2 s}\over {\partial \phi ^2}} = -2s.
\label{sder2}
\ee
These formulae illustrate that the function $s$ is perfectly suitable for describing the effects of rotation (rotation-induced symmetry) in magnetic coordinates. 

\noindent
For the sake of simplicity, we shall ignore all general-relativistic corrections (including the important effect of {\it frame dragging}) throughout this paper. Since we are not going to discuss the polar cap electrodynamics, all these corrections may only unnecessarily complicate the resulting formulae.     

\noindent
We assume, that the magnetic field in the innermost magnetosphere (well  within the LC) can be presented as 
\be
{\bf B} = {\bf B}^{\rm (d)} + \epsilon ~{\bf B}^{\rm (1)},
\label{B1-1}
\ee
where 
\be 
{\bf B}^{\rm d} = {{B_0^{\rm d}}\over {\eta ^3}}~\left(\cos \theta {\bf e}_{\rm r} + {1\over 2} \sin \theta {\bf e}_{\theta }\right)
\label{Bd}
\ee
is a pure dipole magnetic field anchored into the NS and
\be
{\bf B}^{\rm (1)} = {\bf B}^{\rm (1)}(\eta, s, \theta, \phi)
\label{B1-2}
\ee
is the magnetic field generated by the self-consistent electric currents in the domain of the magnetosphere with open field lines.  The parameter $\epsilon \lsim 1$ determines the strength of the ${\bf B}^{\rm (1)}$-component relative to the dipole one and also constrains the limiting value of the density of primary (electron) current at the polar cap surface (see condition [\ref{eps-2}] below). 

The ${\bf B}^{\rm (1)}$-component and the corresponding density of the self-consistent electric current are determined by the Maxwell's equations (that, within the notations, coincide with equations [24]-[26] of MH05)
\be
\epsilon ~\nabla \times {\bf B}^{\rm (1)} = {{4~\pi }\over c}~{\bf j}^{\rm (1)},
\label{curlB1}
\ee
\be
\nabla \cdot {\bf B}^{\rm (1)} = 0,
\label{divB1}
\ee
\be
\nabla \cdot {\bf j}^{\rm (1)} = 0.
\label{divj1}
\ee
Equations (\ref{curlB1})-(\ref{divj1}) are the standard magnetostatic equations. Before we proceed with the analytic solution to these equations, we should specify the fundamental requirements to the expected solution. First, we will be solving equations (\ref{curlB1})-(\ref{divj1}) in magnetic spherical polar coordinates $(r, \theta, \phi)$. Second, in our previous study (MH05) we have demonstrated that the generic current of primary electrons relativistically flowing along the dipole magnetic field of a NS is capable of generating the perturbation to the dipole magnetic field that has a $1/r^2$-radial dependence. The fact that this current of primary electrons generates the $1/r^2$ correction to the magnetic field and matches the self-consistent current we derive in this paper at low altitudes (and in a small-angle approximation) is very suggestive. Also, the leading term in the charge density (all the way from the polar cap surface to the LC)  scales as $1/r^3$ which means that  from basic dimensionality analysis $B^{(1)}$ should scale as $r\rho \sim 1/r^2$. 
So, we will be searching for the solution to equations (\ref{curlB1})-(\ref{divj1}) having the same radial dependence.  Third, both the magnetic field and density of electric current are expected to depend on angular variables through the function $s$ and its derivatives over $\theta $ and $\phi $. This is important because in magnetic coordinates it is the function $s$ that can be used in an elegant way to describe the effect of rotation and symmetry with respect to the rotation axis, and hence to ensure that the corresponding solution to Maxwell's equations (\ref{curlB1}) - (\ref{divj1}) will posses rotational symmetry.  Finally, the $\Omega $-dependence will enter the solution (see also MH05) through the dimensionless radial coordinate, $x = r/R_{\rm lc} \equiv \Omega r/c$.  Physically, this means that we choose to express the solution for ${\bf B}^{(1)}$ as $(\Omega r/c) (B_0^{\rm d}/\eta ^3) {\bf X}$, where $\bf X$ is some vector function (to be solved) that depends only on the function $s$ and its derivatives over $\theta $ and $\phi $. Likewise, the current density in equation (\ref{curlB1}) becomes automatically proportional to the $\Omega B_0^{\rm d}/2\pi \eta ^3$, which is just the amplitude of the current density of primary electrons. This general reasoning is   sufficient to construct the exact solution for ${\bf B}^{(1)}$ and ${\bf j}^{(1)}$ (note that ${\bf j}^{(1)}$ is just one of the three terms contributing to ${\bf j}$, as discussed right after equation [\ref{jtot}]) having the well-specified bulk properties. Thus, what remains to be determined is the exact form of angular dependence for our solution. 

\section{Analytic Solution}

The general solution for ${\bf B}^{\rm (1)}$ satisfying solenoidality condition (\ref{divB1}) and scaling as $1/x^2$ (and therefore can be dubbed as the 3D-monopole-type solution) reads 
\be 
B_{\rm r}^{\rm (1)} = {{B_0^{\rm d}}\over {\eta _{\rm lc}^3}} {{f^{\rm p}(s)}\over {x^2}}, 
\label{B1r}
\ee
\be
B_{\theta }^{\rm (1)} = {{B_0^{\rm d}}\over {\eta _{\rm lc}^3}} {1\over {x^2}}~~\left( g^{\rm p}(s)~{{\partial s}\over {\partial \theta }} + h^{\rm t}(s)
~{1\over {\sin \theta }} {{\partial s}\over {\partial \phi }} \right),
\label{B1theta}
\ee
\be
B_{\phi }^{\rm (1)} = {{B_0^{\rm d}}\over {\eta _{\rm lc}^3}} {1\over {x^2}}~~\left( g^{\rm p}(s)~{1\over {\sin \theta }}~{{\partial s}\over {\partial \phi }} - 
h^{\rm t}(s)~{{\partial s}\over {\partial \theta }} \right),
\label{B1phi}
\ee
where 
\be
g^{\rm p}(s) = {1\over {1-s^2}}
\label{g^p}
\ee
is uniquely determined by the solenoidality of ${\bf B}^{\rm (1)}$.  Although this is a rigorous result, the singularity on the symmetry axis ($s = 1$) should be excluded from the solution domain for the simple physical reason that the symmetry axis is a current-free area (see also discussion following equation [\ref{j1phi-2}] below).  In equations (\ref{B1r})-(\ref{B1phi}) $x = \eta /\eta _{\rm lc}$,  $\eta = r/R$ and $\eta _{\rm lc} = R_{\rm lc}/R$.
 
\noindent
Similarly, we can write the general solution for ${\bf j}^{\rm (1)}$ satisfying (\ref{divj1}) as
\be 
j_{\rm r}^{\rm (1)} = {{i_0}\over {x^3}}~p(s), 
\label{j1r}
\ee
\be
j_{\theta }^{\rm (1)} = {{i_0}\over {x^3}}~\left( q^{\rm p}(s)~{{\partial s}\over {\partial \theta }} + q^{\rm t}(s)
~{1\over {\sin \theta }} {{\partial s}\over {\partial \phi }} \right),
\label{j1theta}
\ee
\be
j_{\phi }^{\rm (1)} = {{i_0}\over {x^3}}~\left( q^{\rm p}(s)~{1\over {\sin \theta }}~{{\partial s}\over {\partial \phi }} - 
q^{\rm t}(s)~{{\partial s}\over {\partial \theta }} \right),
\label{j1phi}
\ee
where 
\be
p(s) = {\partial \over {\partial s}}~\left[ (1-s^2)~q^{\rm p}(s) \right],
\label{p}
\ee
as a direct consequence of the solenoidality of current (see [\ref{divj1}]).

\noindent
By comparing the LHS of equation (\ref{curlB1}) with formulae (\ref{j1r})-(\ref{j1phi}), we find that 
\be
i_0 = - \epsilon {{\Omega~B_0^{\rm d}}\over {4\pi \eta _{\rm lc}^3}} ,
\label{i0}
\ee
\be
q^{\rm p} = h^{\rm t},
\label{q^p}
\ee
\be
q^{\rm t} = - \left(  {{\partial f^{\rm p}}\over {\partial s}} + g^{\rm p} \right) .
\label{q^t}
\ee 
Now we can present ${\bf j}^{\rm (1)}$ as a superposition of poloidal (superscript ``$\rm p$") and toroidal (superscript ``$\rm t$") components,
\be
{\bf j}^{\rm (1)} = {\bf j}^{\rm p} + {\bf j}^{\rm t}
\label{j1-2}
\ee
and write
\be
{\bf j}^{\rm p} = \alpha ^{\rm p}~{\bf B}^{\rm p},
\label{j^p-2}
\ee
\be
{\bf j}^{\rm t} = \alpha ^{\rm t}~{\bf B}^{\rm t},
\label{j^t-2}
\ee
where ${\bf B}^{\rm p}$ and ${\bf B}^{\rm t}$ are the poloidal and toroidal components of ${\bf B}^{\rm (1)}$, and $\alpha ^{\rm p}$, $\alpha ^{\rm t}$ are some scalar functions that will be determined below. 

\noindent 
By substituting expressions (\ref{i0}) - (\ref{q^t}) into formulae (\ref{j1r}) - (\ref{j1phi}) and making use of equations (\ref{curlB1}) and (\ref{j1-2}) - (\ref{j^t-2}), we get 
\be
f^{\rm p} = {s\over {1-s^2}},
\label{f^p}
\ee
\be
q^{\rm p} = h^{\rm t} = {1\over {(1-s^2)^{3/2}}},
\label{q^p-2}
\ee
\be
q^{\rm t} = - {2\over {(1-s^2)^2}},
\label{q^t-2}
\ee
\be
\alpha ^{\rm p} = - {{\epsilon ~\Omega }\over {4~\pi~x~\sqrt{1-s^2}}},
\label{alpha^p}
\ee
\be
\alpha ^{\rm t} = {{\epsilon ~\Omega }\over {2~\pi~x~\sqrt{1-s^2}}}.
\label{alpha^t}
\ee
Thus, the components of the magnetic field (\ref{B1-1}) and electric current density (\ref{j1r})-(\ref{j1phi}) can be presented as
\be
B_{\rm r} = {{B_0^{\rm d}}\over {\eta ^3}}~\left(  \cos \theta + \epsilon ~x~{s\over {1-s^2}}  \right) ,
\label{Br}
\ee
\be
B_{\theta } = {{B_0^{\rm d}}\over {\eta ^3}}~\left[  {1\over 2}~\sin \theta + {{\epsilon ~x}\over {1-s^2}}~\left( {{\partial s}\over {\partial \theta }} + 
{1\over {\sqrt{1-s^2}}}~{1\over {\sin \theta }}~{{\partial s}\over {\partial \phi }}  \right) \right] ,
\label{Btheta}
\ee
\be
B_{\phi } = {{B_0^{\rm d}}\over {\eta ^3}}~{{\epsilon ~x}\over {1-s^2}}~\left( {1\over {\sin \theta }}~{{\partial s}\over {\partial \phi }} - 
{1\over {\sqrt{1-s^2}}}~{{\partial s}\over {\partial \theta }}  \right) ,
\label{Bphi}
\ee
and
\be
j_{\rm r}^{\rm (1)} = - \epsilon ~{{\Omega B_0^{\rm d}}\over {4\pi \eta ^3}}~{s\over {(1-s^2)^{3/2}}},
\label{j1r-2}
\ee
\be
j_{\theta }^{\rm (1)} = - \epsilon ~{{\Omega B_0^{\rm d}}\over {4\pi \eta ^3}}~{1 \over {(1-s^2)^{3/2}}}~\left( {{\partial s}\over {\partial \theta }} - {{2}\over {\sqrt{1-s^2}}}~{1\over {\sin \theta }}~{{\partial s}\over {\partial \phi }} \right) ,
\label{j1theta-2}
\ee
\be
j_{\phi }^{\rm (1)} = - \epsilon ~{{\Omega B_0^{\rm d}}\over {4\pi \eta ^3}}~{1 \over {(1-s^2)^{3/2}}}~\left( {1\over {\sin \theta }}~{{\partial s}\over {\partial \phi }} + {{2}\over {\sqrt{1-s^2}}}~{{\partial s}\over {\partial \theta }} \right).
\label{j1phi-2}
\ee
One can see that formulae (\ref{Br})-(\ref{j1phi-2}) have a singularity at the rotation axis, $s = 1$. The physical meaning of this singularity can be better understood by evaluating the GJ charge  density (see formula [\ref{GJ1}] and discussion following formula [\ref{GJ2}] below) at the polar cap surface:  the surface value of the GJ charge density decreases very near the rotation axis. This means that the maximum charge density of electrons that can be ejected from the polar cap surface should also decrease toward the rotation axis. It is important that, on the magnetic field lines surrounding the rotation axis, the electrostatic potential is constant, and therefore the accelerating electric field should vanish on these field lines. Thus, according to our solution, the polar cap region near the rotation axis is a current-free region that is also unfavorable for particle acceleration. Apparently, our analytic approach is not applicable to this region, and the field lines whose footpoints have $s = s_0 = 1$ should be excluded from consideration. We should also point out that, although our solution still implies some enhancement of the magnetic energy density towards the rotation axis (collimation of the field lines towards the rotation axis), the magnetic flux through the surface normal to the open field lines is constant and is by no means singular even at $s \rightarrow 1$.  

\noindent
Now we should add to ${\bf j}^{(1)}$ the (poloidal) current density of primary electrons,
\be
{\bf j}_{\rm prim}^{p} =   - {{\Omega _0 B_0^{\rm d}}\over {2\pi \eta ^3}}~\Lambda _0 
\left( \cos \theta {\bf e}_{\rm r} + {1\over 2} \sin \theta {\bf e}_{\theta } \right),
\label{je}
\ee
where 
\be
\Lambda _0 \approx \cos \chi  + {3\over 2}\theta \sin \chi \cos \phi ,
\label{Lambda0}
\ee
in a small-angle approximation ($\theta \ll 1$) that is valid at small altitudes above the pulsar polar cap surface. Note that, in MH05 the solution for ${\bf B}^{(1)}$ is determined by the density of primary electron current (equation [\ref{je}]) only. In this paper we find that, in the pair-starved regime, there is a new type of the self-consistent solution that implies a gradual build-up of additional current (on the top of primary electron current) at increasingly higher altitudes while preserving the condition for the occurrence of non-vanishing accelerating electric field, ${\bf E}_{\parallel } < 0$.  The magnitude of ${\bf B}^{(1)}$ given by formulae (\ref{Br})-(\ref{Bphi}) is significantly larger than that of the corresponding component calculated in MH05.       

\noindent
The total density of poloidal current can now be written as
\be
{\bf j}_{\rm tot}^{\rm p} = {\bf j}_{\rm prim}^{\rm p} + {\bf j}^{\rm p}.
\label{JP}
\ee
Note also that formulae (\ref{alpha^p}), (\ref{alpha^t}) imply that in the region (still within the LC) where ${\bf B}^{\rm (1)}$  dominates over ${\bf B}^{\rm (d)}$, the magnetic surfaces (that do not intersect) are described by equation
\be
x~\sqrt{1-s^2} = const.
\label{f-line}
\ee
On each magnetic surface, the angular velocity, $\Omega $, should satisfy the following condition (also known as the Ferraro's isorotation law), 
\be
{\bf B}\cdot \nabla \Omega  \equiv {\bf B}^{\rm p}\cdot \nabla \Omega  = 0,
\label{BnablaOmega}
\ee
meaning that $\Omega $ is a function of $w=x~\sqrt{1-s^2}$ only (in other words, the angular velocity is constant along each poloidal field line), thus ensuring that there is no winding-up of the magnetic field lines. One can easily verify this by direct substitution. Formula (\ref{BnablaOmega}) refers to the component ${\bf B}^{(1)}$ or to the domain where ${\bf B}^{(1)}$ dominates over ${\bf B}^{\rm d}$. In the region where the magnetic field is mostly dipolar, the angular velocity of open field lines has two choices: (1) be constant everywhere (solid-body rotation); and (2) be constant along the field lines (in this case $\Omega $ will be a function of $\sin ^2 \theta /x$ that is a counterpart of function $w$ for a pure dipole field).  However, there is a problem with the second choice, simply because the dipole magnetic field (except the aligned case, $\chi = 0$) is not symmetric with respect to the rotation axis.  Therefore, any differential rotation of the dipole field lines without crossing and entanglement would be topologically and physically prohibited, especially within the light cylinder.  In this study we assume that, in the domain where the dipole field dominates, the angular velocity of open field lines is constant, $\Omega = \Omega _0 = const$ (solid-body rotation), but as ${\bf B}^{(1)}$ begins to dominate $\Omega = $ const is no longer required.   

\noindent
To illustrate the effect of differential rotation of open field lines (in the domain where ${\bf B}^{(1)}$ prevails), we may approximate $\Omega (w)$ by the simple formula 
\be
\Omega  = {\Omega_0 \over {1+aw}} ,
\label{Omega1}
\ee
where $\Omega _0$ is the angular velocity of the NS, and $a$ is a parameter ($\sim 1$) to be determined by matching the solution for $\Omega $ beyond the LC with that within the LC. This formula illustrates the simplest possible way for the angular velocity of open field lines to transit from the solid-body to differential rotation.  It is, by no means, the actual solution for $\Omega $ and will be used below just to examine and clarify some of the properties of our solution. 

\noindent
Because our solution implicitly incorporates differential rotation of surfaces of constant magnetic flux, we should be able to smoothly match the differential angular velocity beyond the LC with the nearly constant angular velocity well inside the LC (see e.g. the corresponding limits of formula [\ref{Omega1}]). It is important to note that in the case of differential rotation given e.g. by equation (\ref{Omega1}), our formalism can be extended even beyond the LC.  More specifically, our formulae (\ref{dB/dt}) - (\ref{drho/dt}) are applicable for subluminal rotational linear velocity, $\Omega r \sqrt{1-s^2} < c$. By using equation (\ref{Omega1}) we can rewrite this fundamental constraint as $w/(1+aw) < 1$ that is actually valid both within the LC ($w \lsim 1$) and beyond the LC ($w > 1$). This means that the problem of superluminal rotation gets automatically fixed: the magnetic field geometry changes to allow the differential rotation of field lines and, at the same time, ensure the subluminal rotation beyond the LC.  

\noindent
For the region well within the LC ($w \ll 1$) , from formula (\ref{Omega1}) we can get, 
\be
\Omega \approx \Omega_0 (1-aw).
\label{Omega2}
\ee

\noindent
Now we can calculate the GJ charge density for the differentially rotating magnetic field lines,
\be
\rho _{_{\rm GJ}} = - {1\over {4\pi }} \nabla \cdot \left(  {\bb}_{\rm rot} \times {\bf B} \right),     
\label{GJ}
\ee
Let us assume that 
\be
\Omega = \Omega _0~F(w),
\label{OmegaF}
\ee
where $F$ is some function of $w$ (e.g. such as determined by formula [\ref{Omega1}]).

By substituting (\ref{OmegaF}) into formula (\ref{GJ}) and performing necessary vector operations we can get
\be
\rho _{_{\rm GJ}} = \rho _{_{\rm GJ}}^0\left( 1 + {1\over 2} {{\partial \ln F}\over {\partial \ln w}}  \right) - \epsilon {{\Omega _0 B_0^{\rm d}}\over {4\pi c\eta ^3}}\left(  {{\partial F}\over {\partial w}} \right) {{x^2}\over {\sqrt{1-s^2}}},
\label{GJ1}
\ee
where
\be
\rho _{_{\rm GJ}}^0 = - {{\Omega B_0^{\rm d}}\over {2\pi c\eta ^3}}~\Lambda (\theta,\phi )
\label{GJ2}
\ee
is the canonical GJ charge density for a pure dipole field (in flat space-time), and   
\be
\Lambda (\theta, \phi) = {1\over 2} \left[ \cos \chi (3\cos ^2 \theta -1) + 3\sin \chi \sin\theta \cos \theta \cos \phi \right].
\label{Lambda}
\ee

\noindent
Note that the second term in equation (\ref{GJ1}) has the opposite sign (the differential velocity is likely to be attributed to the decrease in angular velocity of open field lines compared to the case of a solid-body rotation, i.e. $\partial F/\partial w < 0$) to the first term and is of order of $\epsilon x^2/\sqrt{1-s^2}$.  Consider the region around the rotation axis and the field lines whose footpoints have the values of $s_0$ in the vicinity of $s_0 = 1$. Given formula (\ref{Omega2}), that is perfectly applicable to the situation under discussion (i.e. where the field line geometry is essentially determined by the component ${\bf B}^{(1)}$, and the differential rotation of open field lines becomes possible), and the fact that $w$ should remain constant along the differentially-rotating poloidal field lines (see formula [\ref{f-line}]), the second term in (\ref{GJ1}) should scale as $(1-s^2)^{-3/2}$, so that as $s = s_0 \rightarrow 1$ it will tend to counterbalance the first term. The GJ charge density should significantly diminish towards the rotation axis before it changes sign near the very axis.  Furthermore, the potential drop along those field lines tends to vanish since the potential is a function of $w$, which is analogous to equation (\ref{BnablaOmega}).  This means that the ejection and acceleration of electrons from the small area of the polar cap surface surrounding the rotation axis shuts off, implying that our solution which assumes relativistic electron outflow is not applicable to this area containing the singularity. This is rather instructive: our solution formally contains a singularity but, at the same time, it indicates that near the singularity the initial model assumption (that electrons can be ejected from any point of the polar cap and then accelerated to relativistic velocity) gets violated. Therefore, the solution itself constrains its applicability by forcing the exclusion of the singular region from the solution domain. A non-singular solution could be, in principle, obtained by incorporating the dynamics of particle acceleration from the polar cap surface into the model and by taking into account the feedback from the global structure of fields and currents (including the return current). Obviously, this kind of approach would dramatically complicate the model and prevent an analytic treatment.

\noindent
To estimate the space charge density, $\rho $, we should exploit the ${\rm E}_{\parallel } = 0$ boundary condition at the polar cap surface, or the approximate condition (valid within the accuracy of $\sim \theta _0^2$, where $\theta _0$ is the canonical angular radius of the polar cap) that
\be
\rho \approx \rho _{_{\rm GJ}}~~~~~~~{\rm at}~~~\eta = 1.
\label{zeroE}
\ee

\noindent
Above the polar cap surface the observer ``sees" the following charge density along the magnetic flux tube  
\be
\rho =  - c^{-1}~j_{\rm tot}^{\rm p}.
\label{rho-2}
\ee
By using formulae (\ref{Br})-(\ref{Bphi}), (\ref{JP}), (\ref{zeroE}) and (\ref{rho-2}) the space charge density of relativistic electrons can be presented as 
\be
\rho = - {{\Omega  B_0^{\rm d} }\over {2\pi c \eta ^3}} {{ B_0^{\rm d}}\over {\eta ^3 B}}
\left[ \Lambda _0 +{{\epsilon  x}\over {2(1-s^2)^{3/2}}} \left( \Lambda (1+2\Lambda _0\sqrt{1-s^2}) + {{\epsilon x}\over {1-s^2}}\right)  +\lambda _0\right].    
\label{rho}
\ee
Here $\Lambda _0$ and $\Lambda $ are given by formula (\ref{Lambda0}) and (\ref{Lambda}), respectively; and 
\be
\lambda _0 = - {1\over 2}{{\epsilon \theta _0 ^2}\over {(1-s_0^2)^{3/2}}} \left( \Lambda _0 + 2 \Lambda _0^2 \sqrt{1-s_0^2} + {{\epsilon \theta _0^2}\over {1-s_0^2}}    \right) - {1\over 2} {\it a} \theta _0^2 \sqrt{1-s_0^2}\left( 3\Lambda _0 + {{\epsilon \theta _0^2}\over {1-s_0^2}} \right),
\label{lambda}
\ee
where $s_0 \approx \cos \chi + \theta _0 \xi \sin \chi \cos \phi $, $\xi = \theta (1)/\theta _0$ ($0 \leq \xi \leq 1$), $\theta (1)$ is the magnetic colatitude of the footpoint of an open magnetic field line. 

\noindent
Note that to constrain the model parameter $\epsilon $ we should impose the condition
\be
\epsilon \ll (1-s_0^2)_{\rm min}^{3/2},
\label{eps-2}
\ee 
implying that, at small altitudes above the polar cap surface, the corrections to $\rho $ and $\rho _{_{\rm GJ}}$ that are $\propto \epsilon $ are smaller than $\sim \theta _0^2$ and do not affect the acceleration of primaries. Apparently, $s_0 = 1$ (rotation axis) is a geometrical singularity (see discussion following equation [\ref{j1phi-2}]), and the field lines with $s_0 = 1$ should be excluded from our solution. However, above the NS surface, in the situation where  $s\rightarrow 1$ the field lines get asymptotically (as $r \rightarrow \infty $) collimated along the rotation axis (see equation [\ref{f-line}]). 
The parameter $\epsilon $ can be more accurately constrained by the condition ${\bf E}_{\parallel }=0$ at the surface.     

\noindent
By using equations (\ref{Br}) - (\ref{Bphi}) one can ``visualize" the field lines by taking cuts or projections in different coordinate planes. However, we must admit that, without having a global solution, one cannot reliably define the coordinates of footpoints of the last open field lines (we assume that the last open field lines are determined by the boundary of the pulsar dead zone where ${\bf E}\cdot {\bf B} = 0$ provided that it gets formed). In this paper, for the sake of illustration, we assume that the last open field lines are emanating just from the rim of the canonical polar cap of radius $r_{\rm pc} = \theta _0 R$ which is, strictly speaking, valid only for the aligned case. 

In Figure 1 we presented a projection of the magnetic field lines in the X-Y plane as viewed down the pole for aligned rotator. The footpoints of the field lines have the same magnetic colatitude ($\xi = 0.35$ and $0.4$ for Figure 1a and 1b, respectively) and magnetic azimuthal angles $\phi = 15^{\circ }$,  $30^{\circ }$, ..., $360^{\circ }$.  For smaller values of $\xi $ there will be much stronger collimation of field lines.  However, we caution that the region with very small values of $\xi $ may be free of particles and can be, therefore, excluded from our solution. Note that the unloaded (dipole) magnetic field lines emanating from this region can get deflected towards the rotation axis by `hoop stress' generated by the surrounding loaded field lines. In Figures 2-5 we depict the field line geometry in the meridional plane (cutting through the rotation and magnetic axes) for different values of obliquity angle, $\chi = 0^{\circ }$, $30^{\circ }$, $60^{\circ }$ and $90^{\circ }$ and for the value of $\epsilon = 0.01$. To plot the field lines we used formulae (\ref{Br})-(\ref{Bphi}) for the fixed values of magnetic azimuth, $\phi = 0^{\circ }$ and $\phi = 180^{\circ }$. In Figures 2-5 we used the Cartesian $X-Z$ plane where the positive $Z$-axis is along the magnetic axis, and the rotation axis (in the oblique cases) is pointing to the upper-right corner (in the orthogonal case the rotation axis is along the positive $X$-axis).  The coordinates are scaled by the LC radius, $R_{\rm lc}$. The magnetic colatitudes of the footpoints of all field lines range from $0.05$ through $1.0$ of the half-angular size of the canonical polar cap with the step of $0.05$ (except the orthogonal case for which $\xi = 0.1, 0.2, ..., 1.0$).  From Figures 2-5 one can clearly see the effect of collimation of field lines along the rotation axis. The magnitude of the collimation depends on the parameter $\epsilon $ and $s_0$ (value of $s$ for the field line footpoint): favorably-curved field lines tend to collimate much more than unfavorably-curved ones. For the aligned rotator (see Figure 2) the effect of collimation may be significant for most field lines emanating from the polar cap. Also, in the aligned case, the field lines experience substantial sweepback (see Figure 1a,b).  It is important to note that, although our approximation is valid only within the canonical LC (up to $\approx 0.5-0.8$ of the LC radius), in Figures 2-6 we presented the magnetic field lines up to the cylindrical radius slightly greater than the LC radius (although some field lines close outside the LC, this is unlikely to happen in the global model, because here we have not included the ${\bf E}\times {\bf B}$ drift and rotation effect). In this way, we believe, the behavior of our solution could be better understood, and its deficiencies are more clearly exposed.  Figures 2-5 show that for small values of $s$ the topology of magnetic field lines in the meridional plane ($X-Z$ plane) is very close to that of a dipole. 

Thus, the topology of the magnetic field lines in the proposed pair-starved model differs from that of the force-free model (see most recent calculation by Kalapotharakos \& Contopoulos, 2008) in two major ways: the field lines in the pair-starved model are significantly more strongly collimated along the rotation axis, whereas the last closed field lines and adjacent open lines may remain closed as opposed to the force-free model where these lines may form the``Y" shape structure due to centrifugal force and favor the formation of a current sheet. However, the detailed comparison of field geometries in the pair-starved regime with those in the force-free/MHD regime will be possible only after we present the closed global analytic solution. We expect that the combined effects of bulk rotation and ${\bf E}\times {\bf B}$ drift current (see formulae [60]-[62] and [76]-[78] in MH05) can significantly modify the topology of field lines in the (rotation) equatorial region. This and other issues will be addressed in a separate study aimed at the derivation of global analytic solution matching the inner solution discussed here.             

\section{Discussion and Conclusions} 

We have constructed a relatively simple self-consistent analytic solution for both the magnetic field and electric current in the inner magnetosphere (within the canonical LC) of a NS with arbitrary obliquity and for the pair-starved regime (see MH04). Our main assumption is that the so-called open magnetic field lines of a pulsar are loaded with relativistically outflowing charges. Also, in compliance with the pair-starved regime, we assume that in addition to the primary electron current there is a supply of charges into the open-field region that is sufficient to feed the electric current capable of distorting the global pulsar magnetosphere. In our model, the electric current ($= - c |\rho |$) is equal to the GJ current ($= - c |\rho _{_{\rm GJ}}|$) only at the polar cap surface and varies with altitude (see formulae [\ref{JP}], [\ref{j1r-2}]-[\ref{j1phi-2}]), remaining less than the local GJ current. Our solution is valid within the LC (say, for $x \sqrt{1-s^2} \lsim 0.5$)  and implies the occurrence of non-vanishing accelerating electric field, ${\bf E}_{\parallel } < 0$. The self-consistent current, ${\bf j}^{\rm (1)}$, produces the effect of collimation of field lines along the rotation axis. For the oblique rotator, the effect of collimation is much more pronounced for the favorably curved  ($\cos \phi \geq 0$) than unfavorably curved ($\cos \phi < 0$) field lines. The strongly collimated favorably-curved field lines (for which $s \rightarrow 1$) may not allow the continuous outflow of electrons. Instead, these field lines may be loaded with electrons returning back to the stellar surface. It is also possible that near the rotation axis an acceleration-free zone can be established. For the unfavorably curved field lines the effect of collimation is much weaker but capable of moving the null-surface (defined by the condition $\oo \cdot {\bf B} = 0$) to much higher altitudes or even completely eliminating it.  At high altitudes and near the rotational equator, where $x \sqrt{1-s^2} \sim 1$, the ${\bf E}\times {\bf B}$-drift current becomes important and should be taken into account. Qualitatively, this effect mitigates the collimation by causing the field lines to flare further away from the rotation axis.  Also, this effect plays an important role in current closure and should be incorporated into a truly global solution. In addition, our solution indicates that asymptotically, when ${\bf B}^{\rm (1)}$ becomes dominant, ${\bf E}_{\parallel } \rightarrow 0$, and therefore the regime of a steady-state acceleration may change to the force-free flow. Thus, our solution could be used to model the transition from the pair starved regime of the space-charge-limited flow to the force-free regime. 

The analytic solution we presented in this study can serve as a prototype for more refined global analytic models of the pair-starved regime.  We expect that in a global solution the key unknown parameters, such as e.g. $\epsilon $ will be more reliably constrained and related to the efficiency of particle acceleration.  Such a global solution will explicitly include the second-order effects we neglected in this study. Hence, it would be interesting to explore, in a self-consistent manner, the effect of the bulk rotation and ${\bf E}\times {\bf B}$-current and extend our analytic solution beyond the canonical light cylinder.  Also, it would be interesting to investigate the effect of a possible change from the space-charge-limited to the force-free flow, as one goes from low to high altitudes, respectively (for the same pair-starved pulsar), and address the global current closure.  In the next study, by using our analytic model, we will calculate the accelerating electric field within the domain of open magnetic field lines and update our previous calculations of ${\bf E}_{\parallel }$ at very high altitudes, so that we will be able to explicitly take into account the effects of essentially non-dipolar global magnetic structure on the occurrence and geometry of high-altitude `slot gaps' (Muslimov \& Harding, 2003, 2004).   
 
\acknowledgments  
We acknowledge support from the NASA Astrophysics Theory Program through the Universities Space Research Association.  Also, we thank anonymous referee for the constructive comments that helped to improve the manuscript.

\newpage
           
~
\vskip 0.3cm
\hskip -2.0cm
\includegraphics[width=100mm]{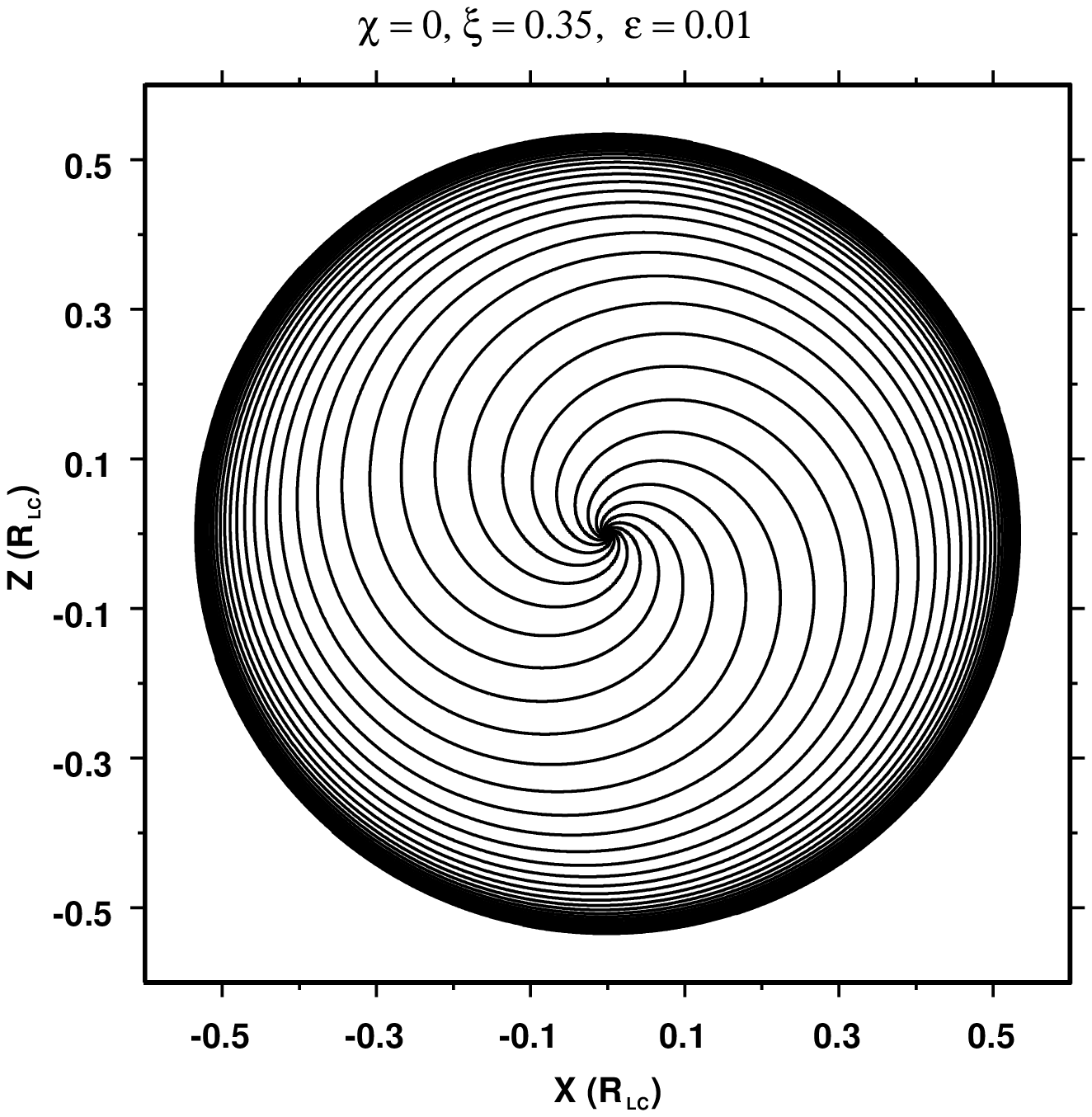}\hskip -1.0cm\includegraphics[width=100mm]{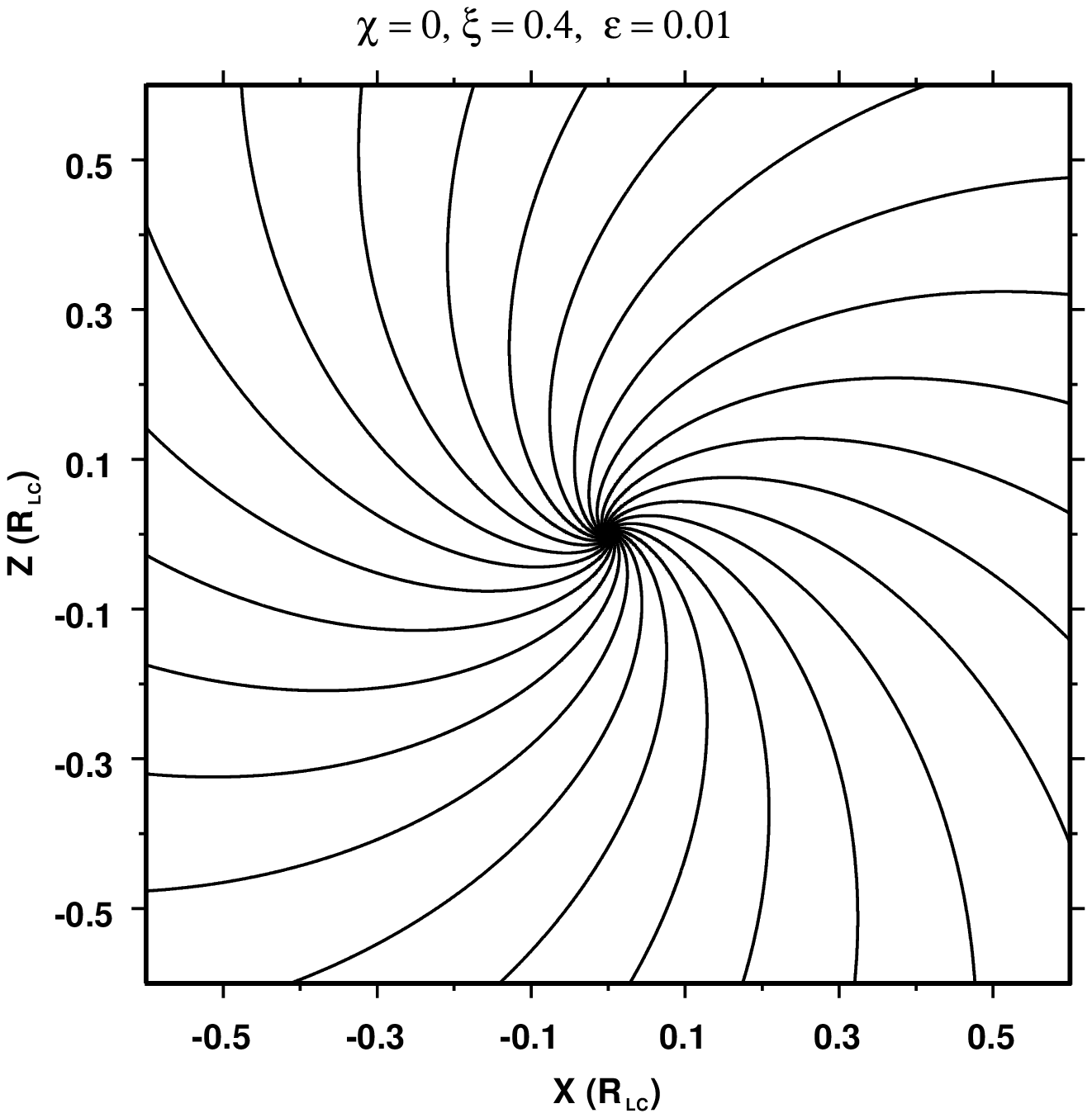}

\figureout{f1a.eps}{
The projections of the magnetic field lines onto the $X-Y$ plane as viewed from the $Z$ axis for the aligned case ($\chi = 0^{\circ}$). The footpoints of the field lines have magnetic colatitude of a) $\xi = 0.35$ (left) and b) $\xi = 0.4$ (right), and azimuths $\phi = 15^{\circ }$,  $30^{\circ }$, ..., $360^{\circ }$.  The parameter $\epsilon = 0.01$.   
    }    

\newpage
\vskip 0.3cm
\hskip -5.0cm
\includegraphics[width=230mm]{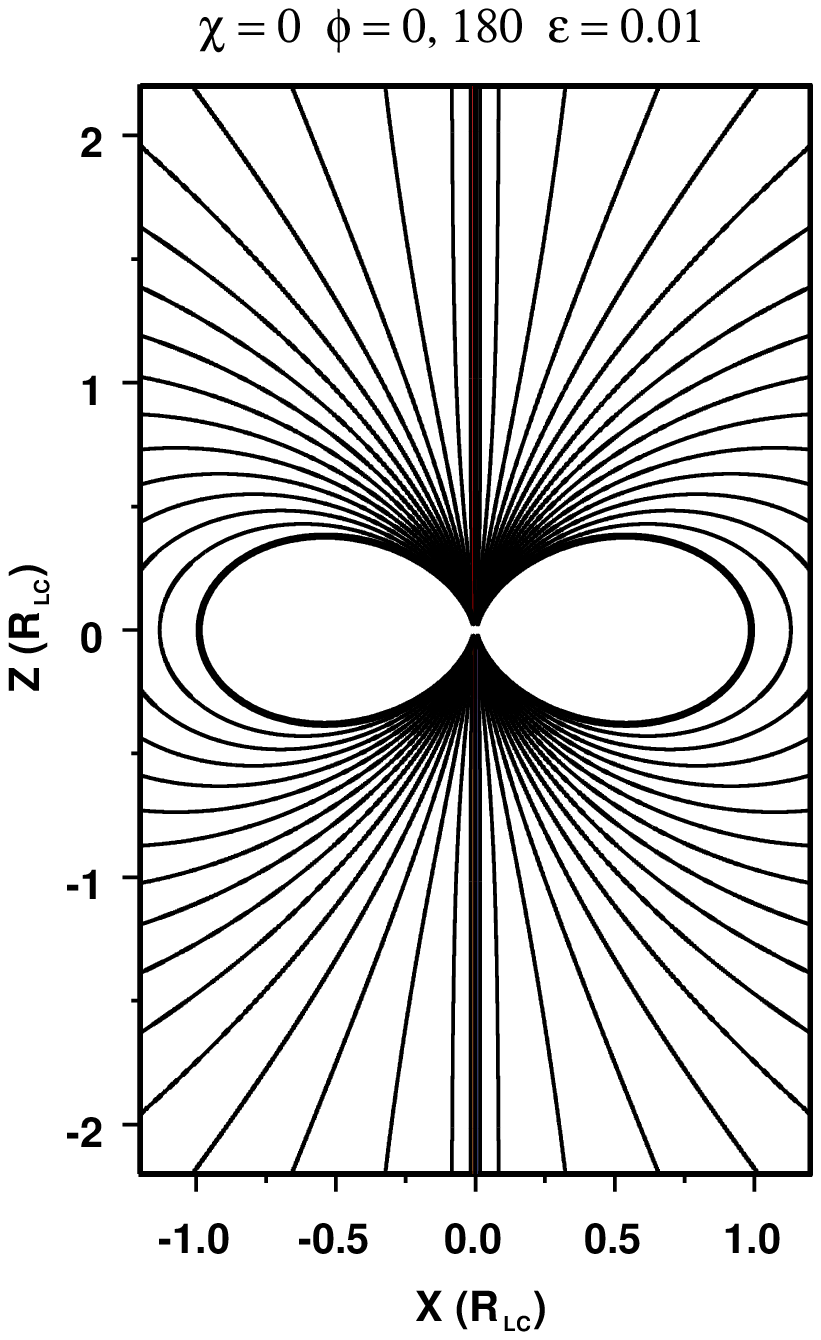}
\figureout{f2.eps}{
The magnetic field lines in the meridional plane for the aligned case ($\chi = 0^{\circ}$). The Cartesian $X$ and $Z$ coordinates are scaled by the canonical value of the LC radius, $R_{\rm lc} = c/\Omega $. The magnetic field lines are emanating from the polar cap and their footpoints have magnetic colatitudes $\xi = 0.05, 0.1, 0.15,...,1.0$. The parameter $\epsilon = 0.01$.   
    }    

\newpage
\hskip -4.0cm
\includegraphics[width=230mm]{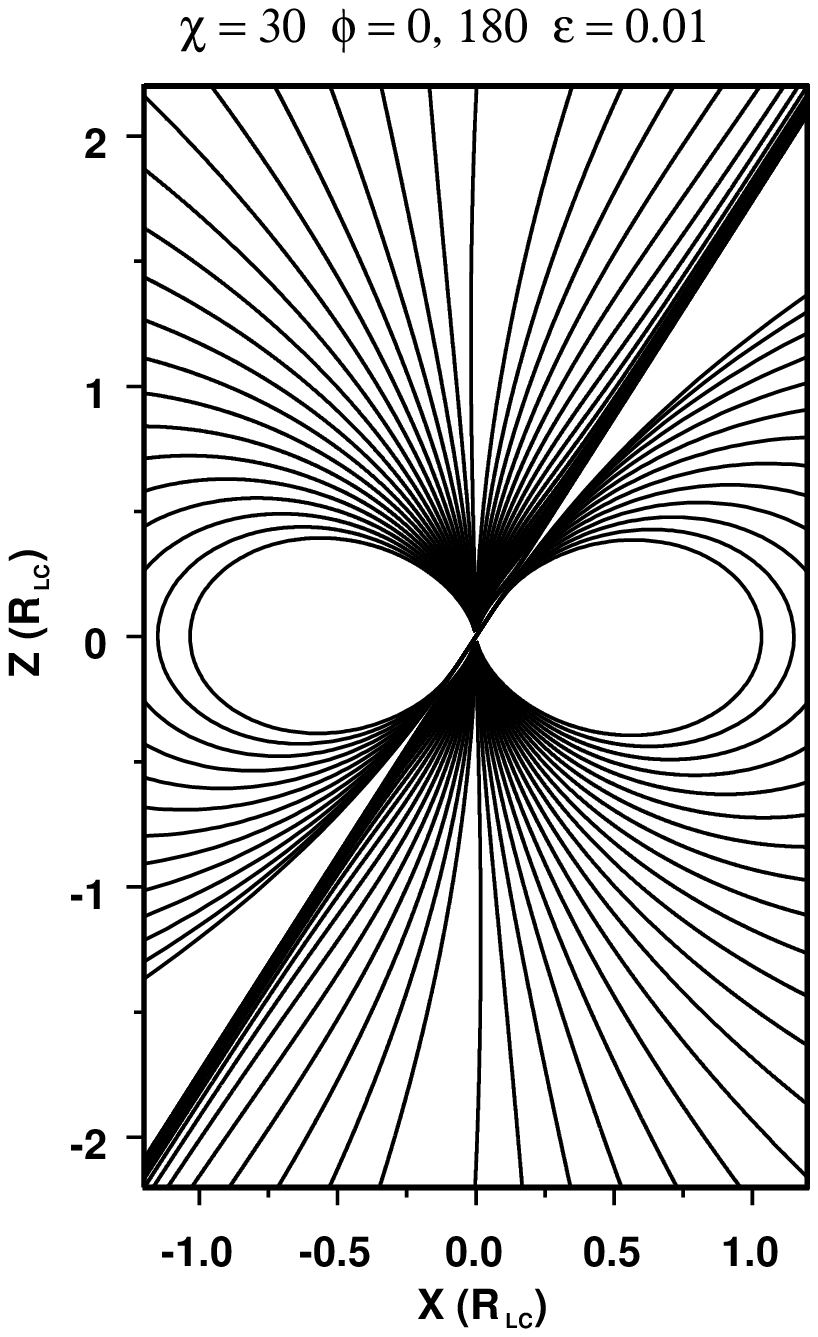}
\figureout{f3.eps}{
The magnetic field lines in the meridional plane cutting through the rotation and magnetic axes for the case $\chi = 30^{\circ}$. Other parameters are the same as in Figure 2.   
    }    

\newpage
\hskip -4.0cm
\includegraphics[width=230mm]{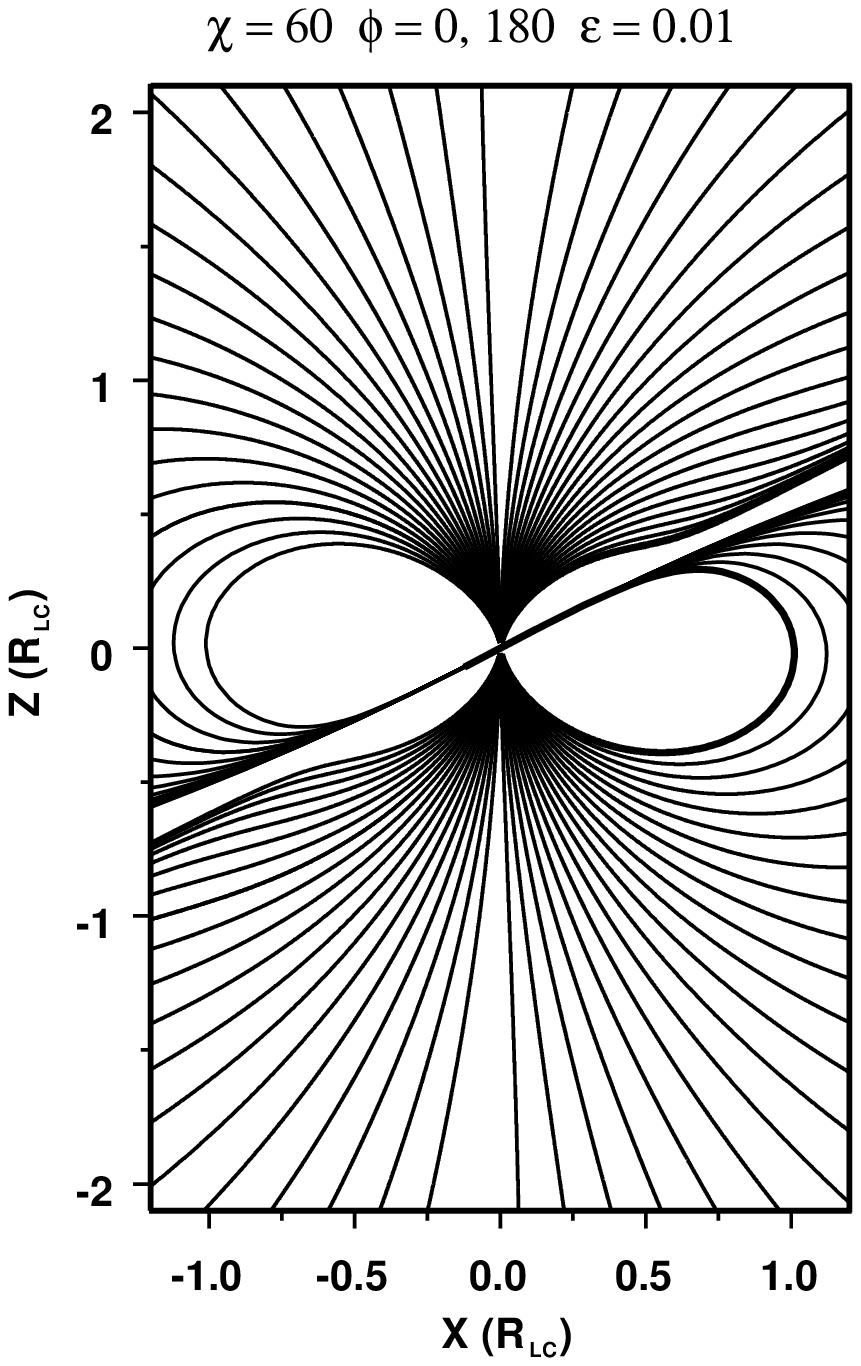}
\figureout{f4.eps}{
 The magnetic field lines in the meridional plane cutting through the rotation and magnetic axes for the case $\chi = 60^{\circ}$. Other parameters are the same as in Figure 2.   
    }    

\newpage
\hskip -4.0cm
\includegraphics[width=230mm]{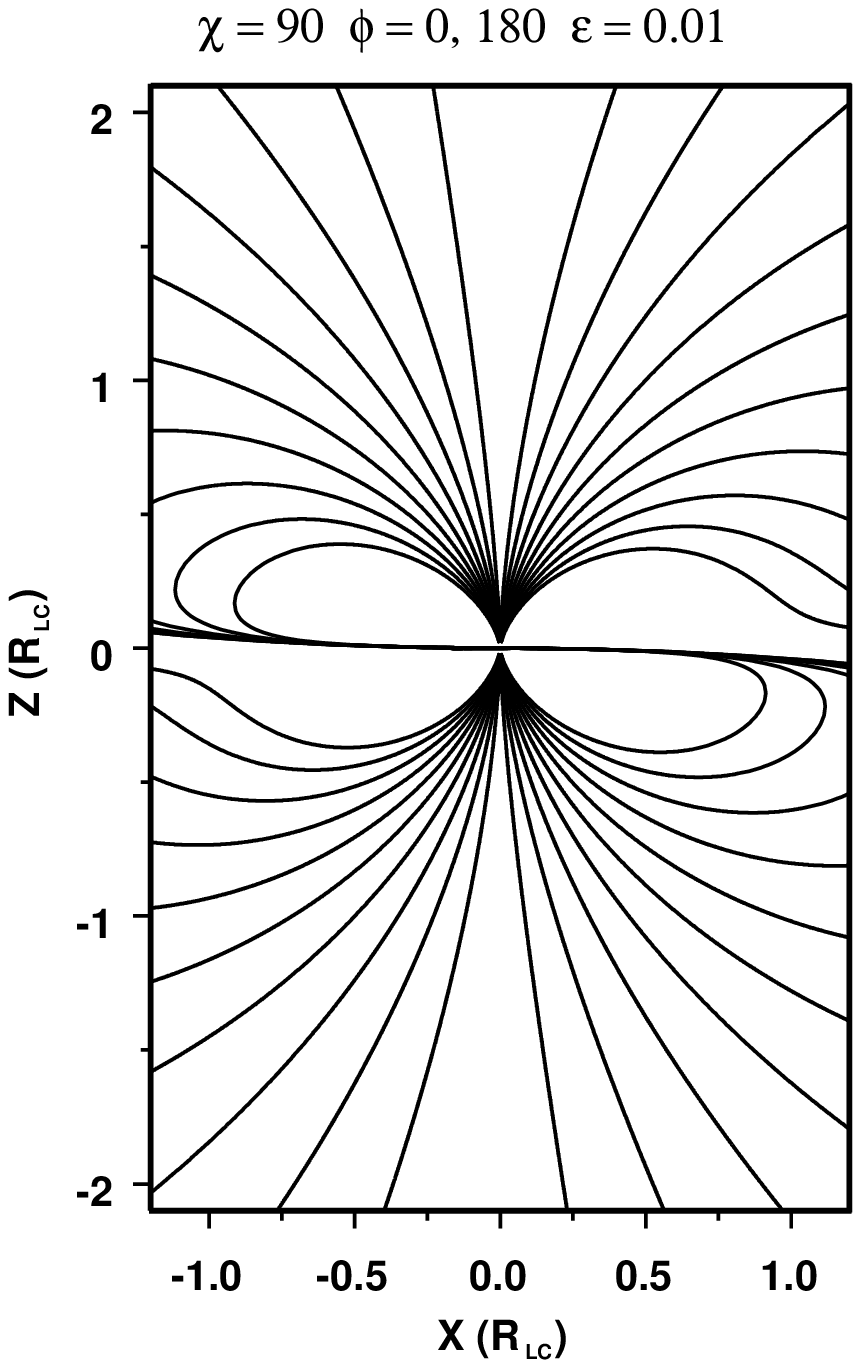}
\figureout{f5.eps}{
 The magnetic field lines in the meridional plane cutting through the rotation and magnetic axes for the orthogonal case ($\chi = 90^{\circ}$). The footpoints of the field lines have magnetic colatitudes $\xi = 0.1, 0.2,...,1.0$, and the parameter $\epsilon = 0.01$.      
    }    

\end{document}